\documentclass[english,aps,prl,twocolumn,nofootinbib,preprintnumbers,groupedaddress,10pt]{revtex4-1}

\usepackage[latin1]{inputenc}
\usepackage{graphicx}
\usepackage{amsmath,amsthm,amssymb}

\usepackage{dcolumn}
\usepackage{hyperref}
\usepackage{bm}
\usepackage{color}

\begin{document}
\preprint{IPMU12-0143}

\title{Strict Limit on CPT Violation from Polarization of 
$\gamma$-Ray Bursts
}

\author{Kenji Toma$^1$, Shinji Mukohyama$^2$, Daisuke Yonetoku$^3$, Toshio Murakami$^3$, Shuichi Gunji$^4$, Tatehiro Mihara$^5$, Yoshiyuki Morihara$^3$, Tomonori Sakashita$^3$, Takuya Takahashi$^3$, Yudai Wakashima$^3$, Hajime Yonemochi$^3$, Noriyuki Toukairin$^4$}

\affiliation{%
${}^1$Department of Earth and Space Science, Osaka University, Toyonaka 560-0043, Japan \\
${}^2$Kavli Institute for the Physics and Mathematics of the Universe, Todai Institutes for Advanced Study, University of Tokyo, 5-1-5 Kashiwanoha, Kashiwa, Chiba 277-8583, Japan\\
${}^3$College of Science and Engineering, School of Mathematics and Physics, Kanazawa University, Kakuma, Kanazawa, Ishikawa 920-1192, Japan\\
${}^4$Department of Physics, Faculty of Science, Yamagata University, 1-4-12, Koshirakawa, Yamagata, Yamagata 990-8560, Japan\\
${}^5$Cosmic Radiation Laboratory, RIKEN, 2-1, Hirosawa, Wako City, Saitama 351-0198, Japan
}

\begin{abstract}
 We report the strictest observational verification of CPT invariance in
 the photon sector, as a result of $\gamma$-ray polarization measurement
 of distant $\gamma$-ray bursts (GRBs), which are brightest stellar-size
 explosions in the universe. We detected $\gamma$-ray polarization
 of three GRBs with high significance levels, and the source distances may be
 constrained by a well-known luminosity indicator for GRBs.
 For the Lorentz- and CPT-violating
 dispersion relation $E_{\pm}^2=p^2\pm 2\xi p^3/M_{Pl}$, where $\pm$
 denotes different circular polarization states of the photon, the
 parameter $\xi$ is constrained as $|\xi|<O(10^{-15})$. Barring precise
 cancellation between quantum gravity effects and dark energy effects,
 the stringent limit on the CPT-violating effect leads to the
 expectation that quantum gravity presumably respects the CPT
 invariance. 
\end{abstract}

\date{\today}

\maketitle

{\bf Introduction.---} 
Lorentz invariance is the fundamental symmetry of Einstein's theory of
relativity. However, in quantum gravity such as superstring theory
\cite{kostelecky89douglas01}, loop quantum gravity
\cite{gambini99alfaro02} and Ho\v{r}ava-Lifshitz gravity
\cite{horava09mukohyama10}, Lorentz invariance may be broken either
spontaneously or explicitly. Dark energy, if it is a rolling scalar
field, may also break Lorentz invariance spontaneously. In the absence
of Lorentz invariance, the CPT theorem in quantum field theory does not
hold, and thus CPT invariance, if needed, should be imposed as an
additional assumption. Hence, tests of Lorentz invariance and those of 
CPT invariance can independently deepen our understanding of the nature
of spacetime.

If CPT invariance is broken then group velocities of photons with
right-handed and left-handed circular polarizations should differ
slightly, leading to birefringence and a phase rotation of linear
polarization. Therefore a test of CPT invariance violation can be
performed with the polarization observations, especially in high
frequency $\gamma$-rays.

The purpose of this letter is to report the strictest observational
verification of CPT invariance in the photon sector, as a result of
$\gamma$-ray polarization measurement of prompt emission of 
distant $\gamma$-ray bursts (GRBs), which are bright stellar-size
explosions in the universe.
We detected $\gamma$-ray polarization of three GRBs with
high significance levels, and we can estimate lower limits on
the source distances for those bursts by a well-known luminosity
indicator.
For the Lorentz- and CPT-violating dispersion relation
$E_{\pm}^2=p^2 \pm 2\xi p^3/M_{Pl}$, where $\pm$ denotes different
circular polarization states of the photon, the parameter $\xi$ is
strictly constrained. The data of one of those bursts, GRB 110721A,
give us the strictest limit, $|\xi|<O(10^{-15})$.
This is the strictest limit on the
CPT invariance violation posed by directly observing the photon sector,
and it is about 8 orders better than the previous limit $|\xi|<10^{-7}$
\cite{fan07}. (As explained later in the present paper, we refute a more
recent limit claimed in \cite{laurent11stecker11}.) Barring precise cancellation 
between quantum gravity effects and dark energy effects, the stringent
limit on the CPT-violating effect leads to the expectation that quantum
gravity presumably respects the CPT invariance.

{\bf Observation and analysis.---}
IKAROS (Interplanetary Kite-craft Accelerated by Radiation Of the Sun)
is a small solar-power-sail demonstrator \cite{kawaguchi08mori10}, and
it was successfully launched on 21 May 2010. IKAROS has a large
polyimide membrane of $20\;$m in diameter, and this translates the solar
radiation pressure to the thrust of the spacecraft. Since the deployment
of the sail on 9 June 2010, IKAROS started solar-sailing towards
Venus. Gamma-ray burst Polarimeter (GAP) \cite{yonetoku11murakami10}
onboard IKAROS is fully designed to measure linear polarization in
prompt emission of GRBs in the energy range of $70-300\;$keV.
Its detection principle is due to anisotropy of Compton scattered
photons. If incident $\gamma$-rays are linearly polarized, the
azimuthal distribution function of scattered photons should
basically shape as $\sin^2\phi$. The GAP consists of a central plastic
scatterer of $17\;$cm in diameter and $6\;$cm in thickness and
surrounding 12 CsI(Tl) scintillators. Coincidence events within
gate time of $5\;\mu$sec between the signal from any CsI and that from
the plastic scintillator are selected for polarization analysis. The
GAP's high axial symmetry in shape and high gain uniformity are keys
for reliable measurement of polarization and avoiding fake modulation
due to background $\gamma$-rays.

The GAP detected GRB~110721A on 21 July 2011 at 04:47:38.9 (UT) at about
0.70 AU apart from Earth. The burst was also detected by Gamma-Ray Burst
Monitor (GBM) \cite{tierney11} and Large Area Telescope (LAT)
\cite{vasileiou11} aboard the Fermi satellite.
Energy fluence of this burst is 
$(3.52 \pm 0.03)\times 10^{-5}\; {\rm erg}\;{\rm cm}^{-2}$ in
$10-1000\;$keV band,
and the photon number flux is well fitted by a power-law 
$N_{\nu} \propto \nu^{\alpha}$ with $\alpha=-0.94 \pm 0.02$ 
in the GAP energy range $70-300\;$keV \cite{tierney11}.
We performed polarization analysis for the entire
duration of GRB~110721A. We clearly detected polarization signal with
polarization degree of $\Pi= 84_{-28}^{+16}\;\%$ and  polarization angle
of $\phi_p=160 \pm 11\;$degree, and null polarization degree is ruled
out with $3.3\sigma$ confidence level.
The $2\sigma$ lower limit on polarization degree is 
$\Pi>35\;\%$ \cite{yonetoku12}.

IKAROS-GAP also detected $\gamma$-ray polarizations of other two GRBs
with high significance levels; with $\Pi=27 \pm 11\;\%$ for GRB~100826A
\cite{yonetoku11b}, and $\Pi=70 \pm 22\;\%$ for GRB~110301A
\cite{yonetoku12}. The detection significance is $2.9~\sigma$ and
$3.7~\sigma$, respectively. 
The $2\sigma$ lower limit on polarization degrees are
$\Pi>6\;\%$ and $\Pi>31\;\%$, respectively.
Therefore we conclude that the prompt
emission of GRBs is highly polarized.

Several of emission mechanisms (e.g., synchrotron emission) proposed for
GRB prompt emission may produce linear polarization as high as 
$\Pi \sim 60\%$ \cite{rees92,yonetoku12}. Observed polarization degree and angle do
not depend on photon energy $E$ significantly in such emission
mechanisms. In order to support this picture, we separately analyzed the
polarization signals for two energy bands, $70-100\;$keV and
$100-300\;$keV, and actually confirmed that the polarizations of
GRB~110721A in the two bands are consistent within the statistical
errors $(\Pi=71_{-38}^{+29}\;\%$ and $\phi_p=155 \pm 15\;$degree for
$70-100\;$keV and $\Pi=100_{-35}^{+0}\;\%$ and 
$\phi_p=161 \pm 14\;$degree for $100-300\;$keV).

As we explicitly show in the next section, the reliable observation of
gamma-ray linear polarization reported here enables us to obtain a
strict limit on CPT violation. 
In order to do this, source distances of the three GRBs are
required to be estimated, but unfortunately their redshifts are
not measured. Instead we use a well-known distance indicator for GRBs,
the $E_{\rm peak}$--peak luminosity correlation,
$L_p = 10^{52.43\pm 0.33} \times (E_{\rm peak}/355\;{\rm keV})^{1.60 \pm 0.082}\;
{\rm erg}\;{\rm s}^{-1}$, where $E_{\rm peak}$ is the peak energy in the source-frame
$\nu F_{\nu}$ spectrum \citep{yonetoku04yonetoku10}.
Once we measure observer-frame $E_{\rm peak}$ and peak flux
we can calculate a possible redshift.
This correlation equation includes systematic uncertainty caused by the data scatter.
Possible redshifts are then estimated to be $0.45<z<3.12$, $0.71<z<6.84$, and $0.21<z<1.09$
with $2\sigma$ confidence level for GRB~110721A, GRB~100826A,
and GRB~110301A, respectively. Hereafter, we use $2\sigma$ lower limit
values for robust discussions for CPT violation.

Before going into details of the limit on CPT violation, however, let us
briefly mention that there are several other works claiming 
detections of linear polarization with low significance, 
but all of the previous reports are
controversial. Ref.~\cite{coburn03} reported detection of strong
polarization from GRB021206 with the {\it RHESSI} solar
satellite. However, independent authors analyzed the same data, and
failed to detect any polarization signals \cite{rutledge04wigger04}. In
these cases, the data selection criteria for the polarization signal was
remarkably different, and the later two authors used more realistic and
reasonable ones. They concluded that the {\it RHESSI} satellite has less
capability to measure the gamma-ray polarization from GRBs even if one
of the brightest GRBs are observed. Ref.~\cite{kalemci07mcglynn07} reported 
detections of polarization with $\sim 2\sigma$
confidence level from GRB~041219 by {\it INTEGRAL}-SPI, and
\cite{gotz09} reported possible detections of time variable polarization
with {\it INTEGRAL}-IBIS data. However, for example in the figure~3 of
\cite{gotz09}, all of the data are not due to the Poisson statistics and
also completely acceptable to non-polarized model while they insist
detection of linear polarization. This is because the systematic or
instrumental uncertainties for the polarization measurement dominate the
photon statistics in these systems. Moreover, the results of SPI and
IBIS for the brightest pulse of GRB~041219 
appear inconsistent
with each other, i.e., the SPI teams detected strong polarization of
$\Pi =98\pm 33$~\% and $\Pi =63^{+31}_{-30}$~\% with $2\sigma$ 
statistical level \cite{kalemci07mcglynn07}, but the IBIS team reported
a strict upper limit of $\Pi<4$~\% \cite{gotz09}. 
(But it should be mentioned that their results for the other temporal 
intervals are consistent.)
Therefore, the
previous reports of the gamma-ray polarimetry for GRBs are all
controversial and, thus, e.g. the argument for the limit on CPT
violation given by \cite{laurent11stecker11} is still open to questions.

Contrary to those controversial previous reports, the detection of
gamma-ray linear polarization by IKAROS-GAP is fairly reliable and thus
can be used to set a limit on CPT violation.

{\bf Limit on CPT violation.---}
Using these highly polarized $\gamma$-ray photons from the cosmological
distance, we constrain the dimension-5, Lorentz violating (LV) operator
in the photon sector. Hereafter,  
$M_{Pl}=(\hbar c/G)^{1/2}=1.22\times 10^{19}\;$GeV is the Planck mass, 
and we shall adopt the unit with $\hbar =c=1$.

In the effective field theory approach \cite{myers03}, LV effects
suppressed by $E/M_{Pl}$ arise from dimension-5 LV operators. In the
photon sector they manifest as the Lorentz- and CPT-violating dispersion
relation of the form 
\begin{equation}
E_{\pm}^2 = p^2 \pm \frac{2 \xi}{M_{Pl}} p^3,  
 \label{eqn:dispersion-relation}
\end{equation}
where $\pm$ denotes different circular polarization states and $\xi$ is
a dimensionless parameter.

If $\xi \neq 0$, then the dispersion relation
(\ref{eqn:dispersion-relation}) leads to slightly different group 
velocities for different polarization states. Hence, the polarization
vector of a linearly polarized wave rotates during its propagation
\cite{gleiser01}.  The rotation angle in the infinitesimal time interval
$dt$ is $d\theta = (E_+ - E_-)dt/2 \simeq \xi p^2 dt/M_{Pl}$. Substituting
$p=(1+z)k$, $dt=-dz/[(1+z)H]$ and 
$H^2=H_0^2[\Omega_m (1+z)^3+\Omega_\Lambda]$, the rotation 
angle during the propagation from the redshift $z$ to the present is
expressed as 
\begin{equation}
\Delta \theta (k,z) 
 \simeq \xi \frac{k^2 F(z)}{M_{Pl} H_0}, \quad
 F(z) = \int^z_0 
 \frac{(1+z') dz'}{\sqrt{\Omega_m(1+z')^3 + \Omega_\Lambda}}. 
\end{equation}
Here, $k$ is the comoving momentum, $H_0=1.51\times 10^{-42}\;$GeV,
$\Omega_m=0.27$ and $\Omega_\Lambda=0.73$.

If the rotation angle differs by more than $\pi/2$ over a range
of momenta $(E_1<k<E_2)$ in which a certain proportion of the total
number of photons in a signal are included, then the net polarization of
the signal is significantly depleted and cannot be as high as the
observed level. This is the case, unless the momentum dependence of the
intrinsic polarization direction of the source is fine-tuned to cancel
the momentum-dependent rotation of the polarization vector induced by
quantum gravity. Such an accidental cancellation is rather unnatural,
and thus we shall not consider this possibility. Hence, the detection of
highly polarized $\gamma$-ray photons by the GAP implies that
$|\Delta\theta (E_2, z)-\Delta\theta (E_1,z)|\leq\pi/2$. In 
order to obtain an upper bound on $|\xi|$ from this inequality we set
$E_1=E_{\rm min}$ and determine $E_2$ by 
$\int^{E_2}_{E_{\rm min}} E^{\alpha} dE/
\int^{E_{\rm max}}_{E_{\rm min}} E^{\alpha} dE=\Pi$, 
where $\Pi$ is the net polarization degree over the GAP energy range
$E_{\rm min} \leq k \leq E_{\rm max}$ and we have adopted the power law
$\propto k^{\alpha}$ with $\alpha<0$ for the photon number
spectrum. This prescription for $E_{1,2}$ corresponds to an ideal
situation in which the detected signal has $100\%$ of polarization
degree and uniform polarization direction over the range 
$E_{\rm min}\leq k<E_2$ but has no polarization in the range 
$E_2\leq k\leq E_{\rm max}$.  With more realistic momentum-dependences
of the polarization degree and direction, $E_2$ would be higher and
hence the bound on $|\xi|$ would be tighter. Without specifying the
nature of the intrinsic polarization of the source, we adopt the one
that gives the weakest bound on $|\xi|$ among those that do not exhibit
the accidental cancellation mentioned above.

For GRB~110721A, the $2\sigma$ lower limits $\alpha>-0.98$ and
$\Pi>35\;\%$ in the whole energy band ($E_{\rm min}=70\;$keV, 
$E_{\rm max}=300\;$keV) lead to $E_2 \simeq 120\;$keV. Setting
$z>0.45$ in 
$|\Delta\theta (E_2, z)-\Delta\theta (E_{\rm min},z)|\leq\pi/2$, we 
obtain the constraint from GRB 110721A as $|\xi| < 7 \times 10^{-15}$.

More accurate constraints are obtained by requiring that 
$\sqrt{Q^2+U^2}/N > \Pi$, where 
$N = \int^{E_{\rm max}}_{E_{\rm min}} E^{\alpha} dE$, 
$Q = \int^{E_{\rm max}}_{E_{\rm min}} E^{\alpha} \Pi_i \cos(2\Delta\theta(E,z))$, and
$U = \int^{E_{\rm max}}_{E_{\rm min}} E^{\alpha} \Pi_i \sin(2\Delta\theta(E,z))$ with
the intrinsic polarization degree $\Pi_i = 1$. Using $\Pi>0.35$ and
$\alpha>-0.98$, we obtain the constraint from GRB 110721A as
\begin{equation}
|\xi| < 2 \times 10^{-15},
\label{eqn:limit}
\end{equation}
which is tighter than the above rough estimate. Alternatively, we may
assume that the intrinsic polarization degree is not as high as $100\%$
but given by the maximum level in the synchrotron mechanism, i.e.,
$\Pi_i = -\alpha/(-\alpha+2/3)$ with $\alpha=-0.98$. This leads to more
stringent limit $|\xi| < 8 \times 10^{-16}$. Generically speaking, if we
assume a lower intrinsic polarization degree then the bound on $|\xi|$
becomes tighter.

From the other GRBs, we obtain weaker constraints. GRB 100826A has 
$2\sigma$ limits as $\Pi >6\;\%$, $\alpha > -1.41$ \cite{yonetoku11b}, 
and $z>0.71$.
Setting $\Pi_i=1$ (or $\Pi_i = -\alpha/(-\alpha+2/3)$), we obtain the
constraint $|\xi|<2 \times 10^{-14}$ (or $|\xi|<1\times 10^{-14}$).
GRB 110301A has $2\sigma$ limits as $\Pi>31\;\%$, 
$\alpha > -2.8$ \cite{foley11}, and $z>0.21$.
Setting $\Pi_i=1$ (or $\Pi_i = -\alpha/(-\alpha+2/3)$), we obtain the
constraint $|\xi|<2 \times 10^{-14}$ (or $|\xi|<1\times 10^{-14}$).

One may consider a more direct constraint from the difference of the 
polarization angles in the two energy bands for GRB 110721A, say
$\Delta\theta(E=170\;{\rm keV},z)-\Delta\theta(E=80\;{\rm keV},z)
<64\;{\rm degree}$ at $2\sigma$ confidence level. This provides
$|\xi|<2 \times 10^{-15}$. If polarization angles are measured 
more accurately as function of energy for GRBs in future, more 
stringent limit would be obtained.

{\bf Comparison with other limits.---}
Our bound (\ref{eqn:limit}) is the strictest limit on the CPT
invariance posed by directly observing the photon sector, and it is
about 8 orders better than the previous limit $|\xi|<10^{-7}$
\cite{fan07}. (As already explained, we consider the limit claimed in
\cite{laurent11stecker11} unreliable.) The constraint from non-detection of  
Ultra-High-Energy (UHE) photon ($E>10^{19}\;$GeV), $|\xi|<10^{-14}$
\cite{galaverni08}, appears to be closer to our bound. However, the
constraint from UHE photon relies on the assumption that the dimension-5
LV operator in the electron sector is sufficiently suppressed
\cite{maccione08}. On the other hand, the previous bound in \cite{fan07}
and our bound do not depend on such an assumption.

The dimension-5 LV operator in the photon sector induces dimension-3
CPT-odd LV operators in the fermion sector by radiative corrections due
to particle interactions. Assuming supersymmetry
\cite{nibbelink05bolokhov05} above $M_{susy} (>{\rm TeV})$, the
radiatively generated dimension-3  CPT-odd LV operators generically have
coefficients of order $b \simeq M_{susy}^2/M_{Pl}$. Hence, existing
experimental bounds on $b$ can be reinterpreted as bounds on $\xi$. For
example, the bound  $|b|<10^{-27}\;$GeV from Xe/He maser \cite{cane04}
implies $|\xi|<10^{-14}$. Our bound (\ref{eqn:limit}) is slightly
stronger than this. On the other hand, the bound $|b|<10^{-33}\;$GeV
from K/He magnetometer \cite{brown10} corresponds to the stronger bound
$|\xi|<10^{-20}$. Note, however, that these bounds inferred from
radiatively generated dimension-3 CPT-odd LV operators are indirect and
rely on supersymmetry. Our bound (\ref{eqn:limit}), on the contrary,
does not rely on supersymmetry and is direct.

In the effective field theory approach \cite{myers03}, there is only one
operator that leads to a linear energy dependence of the speed of light
in vacuum, and it is the dimension-5 CPT-odd LV operator considered in
the present paper. Constraints on the same operator from observation of
energy dependence of GRB light curves \cite{amerino98} are not as
significant as those from observation of polarization such as ours. For
this reason, once the stringent bound from the latter type of
observation is imposed on the unique dimension-5 LV operator, it is
natural to interpret the former type of observation as limits on the
dimension-6 LV operator. In this case, observation of GRB~090510 by
Fermi satellite \cite{abdo09} leads to the lower bound on the quantum
gravity mass scale as $M_{QG,2} >10^{11}\;$GeV. This is consistent with
natural expectation that the quantum gravity mass scale is of the order
of the Planck mass.

{\bf Conclusion.---}
In some quantum gravity theories such as superstring theory
\cite{kostelecky89douglas01}, loop quantum gravity
\cite{gambini99alfaro02} and Ho\v{r}ava-Lifshitz gravity
\cite{horava09mukohyama10}, Lorentz invariance may be broken either
spontaneously or explicitly. Dark energy, if it is a rolling scalar
field, may also break Lorentz invariance spontaneously. Barring precise
cancellation between quantum gravity effects and dark energy effects, the
stringent limit (\ref{eqn:limit}) on the Lorentz- and CPT-violating
parameter $\xi$ then naturally lead us to the expectation that quantum
gravity theory and/or state may break Lorentz invariance but presumably
respect the CPT invariance. The celebrated CPT theorem in quantum field
theory assumes Lorentz symmetry and locality. In the absence of Lorentz
symmetry, the CPT invariance, if needed, should be imposed as a part of
the definition of the theory. In LV but CPT invariant theories, the
parameter $\xi$ exactly vanishes and thus all existing limits on $\xi$
are trivially satisfied.

~\\~
\begin{acknowledgments}
 We thank the referee for useful comments to improve our manuscript.
 K. Toma and S. Mukohyama are responsible for the theoretical
 discussions of this paper, and D. Yonetoku, T. Murakami and IKAROS-GAP
 team contributed to the observation and data analysis of $\gamma$-ray
 polarization. This work is supported by the Grant-in-Aid for Young
 Scientists (S) No.20674002 (DY), Young Scientists (A) No.18684007 (DY),
 Grant-in-Aid for Scientific Research 24540256 (SM), 21111006 (SM),
 Japan-Russia Research Cooperative Program (SM), and Research
 Fellowships for Young Scientists No.231446 (KT) from the Japan Society
 for the Promotion of Science (JSPS), and also supported by the Steering
 Committee for Space Science at ISAS/JAXA of Japan and the World Premier
 International Research Center Initiative (WPI Initiative), MEXT,
 Japan. 
\end{acknowledgments}


\end{document}